TOWARD LEARNING SOCIETIES FOR DIGITAL AGING 1

# Toward Learning Societies for Digital Aging

Ning An

The Gerontechnology Lab, The Hefei University of Technology

## Author Note

I have no conflicts of interest to disclose. This work was partially supported by the Anhui Province Higher Education Institution Quality Project "Smart Eldercare" (2019kfkc010) and the Hefei University of Technology Teaching Reform Demonstration Course Project" Smart Eldercare" (KCSZ2020038). Correspondence concerning this article should be addressed to Ning An. Email: ning.g.an@acm.org

**Abstract**

The global aging population presents significant challenges for societies worldwide, particularly in an increasingly digitalized era. The Learning Society is crucial in preparing different societies and their people to address these challenges effectively. This paper extends this concept and proposes a new learning ecosystem – Learning Societies for Digital Aging - empowering all members across various sectors from different ages to acquire and develop the necessary knowledge, skills, and competencies to navigate and thrive in an increasingly digital world. It presents seven guiding principles for developing this learning ecosystem: 1) Centering Humanistic Values, 2) Embracing Digital, 3) Cultivating Learning Societies, 4) Advancing Inclusiveness, 5) Taking Holistic Approaches, 6) Encouraging Global Knowledge Sharing, and 7) Fostering Adaptability. By integrating these guiding principles into the design, implementation, and evaluation of formal, nonformal, and informal learning opportunities for people of all ages, stakeholders can contribute to creating and nurturing learning societies that cater to aging populations in the digital world. This paper aims to provide a foundation for further research and action toward building more inclusive, adaptive, and supportive learning environments that address the challenges of digital aging and foster more empathetic, informed, and prepared societies for the future of aging.

**Toward Learning Societies for Digital Aging**

Amid global volatility spanning social, economic, political, educational, and health landscapes, two prominent megatrends persist: the inexorable growth of the aging population and the rapid advancement of digital technologies. The challenge is integrating these two megatrends to yield mutual benefits (Kleinman et al., 2021). As we move towards a digital future, it is crucial to ensure that all age groups, including older adults, are empowered to navigate and thrive in this evolving landscape. This paper posits the new concept of "learning societies for digital aging" that encompasses the creation of inclusive, adaptive, and supportive environments that foster learning and development for individuals of all ages in the context of our digital world. It seeks to explore and elaborate on this concept, with a particular focus on the interplay between aging, digital technology, and learning.

Aging is a universal and dynamic experience shared by fortunate individuals, transcending cultural, social, and economic boundaries. With the rapid digital evolution, a new form of aging has emerged – "digital aging." It refers to the process through which individuals, communities, and societies adapt and respond to the integration of digital technologies into various aspects of life, especially as it pertains to aging and older adults. Digital aging encompasses not only the adoption and use of digital tools and services by older adults but also the broader implications of digitalization for aging populations, including how digital technology can support and enhance well-being, social connectedness, and lifelong learning. By cultivating "Learning Societies for Digital Aging," we can ensure that individuals, throughout their lifespans, continue to develop the necessary knowledge, skills, and competencies to adapt and thrive in an ever-evolving digital environment. We choose the plural "Learning Societies" instead of the singular "Learning Society" for the following two reasons. First, it implies the existence of multiple societies, communities, or contexts working towards the same goal. It suggests we are working on a more diverse and inclusive approach, acknowledging that different societies may have unique needs, challenges, and strategies for fostering learning in the digital age. Second, it conveys the idea that there is no one-size-fits-all solution and that a variety of approaches and strategies can contribute to building learning environments for digital aging across different societies and contexts.

This paper will outline the guiding principles for developing the learning ecosystem of "Learning Societies for Digital Aging," drawing from various disciplines, including gerontology, education, sociology, and information technology. We will then discuss the learning ecosystem, which encompasses formal, nonformal, and informal learning approaches, as well as innovative strategies that can be employed to address the diverse needs of older adults and other population groups in the digital age. Through a range of case studies and examples, we aim to illustrate how these principles can be applied in practice, fostering the development of learning societies that empower individuals to navigate the challenges and opportunities presented by digital aging successfully.

By advancing our understanding of learning societies for digital aging and providing insights on their implementation, this paper aims to contribute to the development of more inclusive, adaptable, and resilient societies capable of supporting the well-being and lifelong learning of individuals across their aging process.

## Guiding Principles

This section will discuss the seven principles that guide the development of our learning ecosystem for learning societies for digital aging. These principles are crucial to understanding and promoting an open and inclusive approach to addressing the challenges and opportunities posed by digital aging.

### Centering Humanistic Values

Centering humanistic values is the first and foremost principle in fostering learning societies for digital aging. It emphasizes the importance of putting people and their needs, experiences, and dignity at the forefront of digital transformation efforts, especially concerning aging populations.

- Prioritizing empathy and compassion: In the design and implementation of digital technologies, we must strive for age-friendly solutions that are accessible, usable, and inclusive for all individuals, regardless of their age, abilities, or preferences.
- Promoting dignity and respect: Treat older adults with dignity and respect, valuing their life experiences, wisdom, and contributions to society.
- Fostering social connectedness: We should leverage digital technologies to enhance social connectedness among older adults by enabling them to maintain and forge new relationships, engage with their communities, and participate in social and cultural activities. This effort can contribute to their well-being and sense of belonging and help them avoid social isolation.

**Embracing Digital**

Promote the adoption and integration of digital technologies in various aspects of aging societies, including healthcare, communication, education, and social services. Highlight the role of technology in enhancing the quality of life for older adults and enabling them to remain active, independent, engaged, and connected.

- Integrating digital tools and services: Seamlessly incorporate digital technologies into various aspects of life, including healthcare, communication, and entertainment.
- Facilitating digital innovation and creativity: Encourage the development of innovative and creative digital solutions to address aging-related challenges.
- Encouraging digital exploration: Build an environment where people feel comfortable and motivated to experiment with digital technologies, explore their benefits, and develop confidence in their digital skills, regardless of age or previous experience.

**Cultivating Learning Societies**

Encourage the development of learning environments that cater to older adults' diverse needs and preferences, including intergenerational and lifelong learning opportunities. Facilitate the sharing of knowledge, experiences, and best practices among different stakeholders.

- Supporting lifelong learning: Encourage individuals to learn, grow, and adapt throughout their lives continuously.
- Encouraging diverse learning opportunities: Provide various learning opportunities to suit different interests, needs, and learning styles.
- Promoting collaborative Learning: Foster environments where individuals can learn from and with each other, sharing experiences, knowledge, and skills.

**Advancing Inclusiveness**

Ultimately, advancing inclusiveness in the context of digital aging is about recognizing and respecting the autonomy and diversity of older adults and creating a supportive and flexible environment that allows them to make their own choices about how to age in a digital society.

- Supporting autonomy and self-determination: Empower older adults to make informed decisions about their engagement with digital technologies. They should be free to choose whether to embrace digital aging and to what degree. For those who opt not to engage with digital technologies, we should support their decision to age gracefully in a digital society.
- Promoting equity and social cohesion: Work towards a more equitable and socially cohesive society for older adults by acknowledging and valuing their life experiences, wisdom, and contributions to society and ensuring that digital technologies do not marginalize or discriminate against them. Moreover, acknowledge and incorporate diverse perspectives from various stakeholders to create inclusive digital solutions.

- Addressing barriers to digital participation: We must try our best to identify and address the various barriers that may prevent older adults from fully participating in digital societies, including affordability, accessibility, and usability of digital technologies, as well as cognitive, physical, and emotional challenges. Sufficiently Addressing these barriers can create a more inclusive and supportive environment for all.

**Taking Holistic Approaches**

Cultivate collaboration among various disciplines, sectors, and stakeholders to create comprehensive, effective, and sustainable solutions. Taking Interdisciplinary Approaches: Encourage interdisciplinary research, education, and practice to foster innovation.

- Encouraging interdisciplinary collaboration: In-depth collaboration across various fields, including gerontology, sociology, psychology, education, policy, and technology, will help develop a comprehensive understanding of the needs and opportunities in digital aging.
- Addressing interrelated factors and needs: Consider the interdependencies among different aspects of life and address the complex, interconnected needs of the aging population.
- Developing scenario-sensitive solutions: Create solutions tailored to the unique needs and circumstances of specific communities.

**Encouraging Global Knowledge Sharing**

Digital aging is a global phenomenon, and we should facilitate international cooperation and the exchange of information, expertise, and resources to address common challenges and capitalize on opportunities collectively.

- Facilitating international cooperation: Promote cooperation and collaboration among countries to address shared challenges related to digital aging.

- Leveraging best practices and lessons learned: Share and apply best practices and lessons learned from different scenarios to inform the development of effective digital solutions.
- Nurturing global networks and partnerships: Build and maintain global networks and partnerships to facilitate the exchange of knowledge, resources, and support.

**Fostering Adaptability**

Recognize the dynamic nature of digital technologies and the changing needs and preferences of aging populations. Foster a culture of continuous learning, innovation, and adaptation to ensure that learning societies remain relevant, effective, and responsive to the evolving needs of an aging population.

- Promoting flexibility and resilience: Encourage individuals, communities, and societies to be flexible and resilient when facing the evolving digital landscape. Anticipate and proactively prepare for the challenges and opportunities that may arise due to digitalization and an aging population.
- Encouraging continuous improvement and innovation: Foster a culture of continuous improvement and innovation, embracing change and learning from experiences.
- Ensuring framework adaptability: Recognize that this learning ecosystem itself should be adaptable to address the evolving nature of digital aging across diverse global contexts. This principle requires continuous evaluation and refinement of the framework's principles and components to better serve the changing needs of individuals and societies in the digital age.

## Digital Empathy

Building upon the guiding principles outlined in the previous section, Digital Empathy emerges as a critical bridge between our foundational values and the practical application within our Learning Ecosystem. It represents a commitment to integrating empathic understanding into all digital interactions, ensuring that our technological advancements resonate with the human experiences of all users, particularly the aging population.

**Defining Digital Empathy**

We define Digital Empathy as the nuanced understanding and facilitation of empathetic interactions within digital environments, spanning a spectrum of relationships among humans, between humans and digital environments, and among digital beings. This philosophical concept profoundly influences the fabric of digital societies, placing human experiences at its core. It underscores the pivotal role Digital Empathy plays in shaping interactions that are not only technologically advanced but also deeply human-centric, making each user's experience integral to the technological advancements.

To further understand this concept, let us explore how Digital Empathy manifests across a spectrum of relationships within digital environments, each offering multifaceted meanings and applications of empathy:

- Among Humans: Digital Empathy enhances communication and understanding between individuals in digital spaces. Fostering a deeper empathic connection makes interactions more meaningful and supportive. This includes the use of technology to convey emotions accurately and sensitively, allowing for richer, more compassionate exchanges that bridge physical distance and cultural divides.
- Between Humans and Digital Environments: This relationship involves interactions with artificial intelligence (AI) and other digital entities capable of understanding and responding to human emotions in a supportive and realistic manner. Digital Empathy here means creating digital assistants, chatbots, and other AI-driven tools that can interpret emotional cues and engage in a way that feels genuinely understanding and caring. The goal is to build trust and provide emotional support through technologies that respect and reflect human emotional needs.

- Among Digital Beings: In this context, Digital Empathy refers to the development of digital entities that can interact with each other in ways that model or simulate empathetic behaviors. This development influences the overall environment, making it more responsive and adaptive. It also enhances user experiences by creating a digital ecosystem that operates under principles of mutual understanding and support. This could involve autonomous systems that communicate with each other to better coordinate responses to human needs or to manage interconnected systems like smart learning systems more effectively.

As we have explored how Digital Empathy manifests in various digital interactions, it is evident that this concept is dynamic, evolving continually over time. The evolution of Digital Empathy from a nascent idea to a future cornerstone of digital interaction reflects broader shifts in society, technology, and our collective values. In the following subsection, we will briefly trace this progression, examining how historical contexts, technological advancements, and changing societal needs have shaped the current understanding of Digital Empathy and hint at its future trajectory.

**The Evolution of Digital Empathy**

Our definition of Digital Empathy has historical roots in the evolution of the concept of empathy. Originally coined in German as "Einfühlung" in 1873, the term initially described the projection of human feelings onto nature and art, highlighting its aesthetic origins (Lanzoni, 2018). The exploration of this concept, however, extends much deeper into human history. The ancient Chinese philosopher Zhuangzi (369BC – 286BC), in his famous "Joy of Fish" story, reflected on empathy with the natural environment. When Zhuangzi observed fish swimming in a river and remarked on their happiness, he suggested that humans could project their feelings and experiences onto the natural world. (Zhuangzi & Brook Ziporyn, 2020). Zhuangzi's empathy toward the fish shows us that empathy need not be confined to human interactions alone.

While the root of empathy went beyond human interactions, psychology later formalized empathy as an essential component of understanding human emotions and relationships, particularly in fields such as psychoanalysis and education. Freud, for instance, considered empathy an intellectual process essential for patient-analyst rapport (Pigman, 1995). In education, historical empathy emerged during the Cold War era (1950-1980) as part of inquiry-based learning initiatives that encouraged students to consider different perspectives (Perrotta & Bohan, 2020). Over time, empathy has evolved from merely an aesthetic theory to a multifaceted tool utilized in the fields of art, psychology, social science, psychiatry, psychotherapy, and neuroscience, where it could function as a technique, a practice, and an aspiration (Lanzoni, 2018).

Researchers often break down empathy into two key dimensions (Decety & Jackson, 2004):

- Affective Empathy: Also known as emotional empathy, affective empathy involves feeling what another person feels, often referred to as emotional contagion. It allows individuals to experience another person's emotions, which can be useful for building deep emotional connections. However, due to its highly emotional nature, it can sometimes lead to biased decision-making.
- Cognitive Empathy: Also known as perspective-taking, cognitive empathy involves understanding what another person is feeling or thinking without necessarily sharing those feelings. It is more aligned with rational thought, allowing individuals to grasp another's emotional state without becoming emotionally overwhelmed. Cognitive empathy helps predict emotional reactions and respond appropriately, making it an essential tool for moral and rational decision-making.

Critiques like Paul Bloom's Against Empathy (2016) focus largely on the emotional dimension of empathy as potentially irrational. This oversimplification misses the importance of cognitive empathy in facilitating reasoned and thoughtful responses to others' emotions.

While scholars dissect these emotional and rational dimensions, science fiction has long foreshadowed how emerging technologies might reshape our empathic capacities. A notable literary example is The Empathy Box from Philip K. Dick's *Do Androids Dream of Electric Sheep?* (Dick, 1968). The Empathy Box serves as a central symbol in the novel, offering a fascinating lens to explore how technology can facilitate and mediate empathetic connections. Although it is fictional, The Empathy Box poignantly anticipates several themes that continue to surface in modern digital environments. First, it illustrates how technology can collapse physical and emotional distance, creating a shared empathic space that transcends individual isolation—an early foreshadowing of today's connected social networks and virtual communities. This effect can be observed in modern digital platforms where large-scale empathetic responses emerge around global humanitarian causes, such as crowdfunding for disaster relief. Second, it raises questions about authenticity and manipulation; by plugging into a collective empathic feedback loop, users risk losing themselves in an experience that may be orchestrated or even deceptive. At the heart of the novel, empathy distinguishes humans from androids—suggesting that while technology might simulate empathic behaviors, it may also blur the line between genuine emotional bonds and artificial replications. Finally, it underscores empathy's evolving role as a measure of "humanity"—an idea echoed today in how AI systems increasingly attempt to read, respond to, or simulate emotional states.

One observable connection between the Empathy Box and today's digital realities is empathy among humans must traverse purely online mediums in many scenarios. Researchers have naturally examined this human layer in digital environments, which often strip away many of the emotional cues essential to face-to-face interactions. Turkle (2011) has explored how digital platforms can both diminish and enhance the capacity for empathy, depending on their structure and use. This nuanced view of digital interaction underscores the importance of designing digital tools that foster empathetic exchanges, even without physical presence, just as The Empathy Box sought to replicate a shared empathic experience. It also demonstrates the need to learn Digital Empathy. Terry and Cain (2016), for instance, raised the concern that the digitization of healthcare may decrease the expression of empathy and advocated introducing digital empathy into healthcare curricula to prepare providers for empathetic interaction via digital platforms. Their advocacy of learning Digital Empathy has opened the door for its application across various professional fields where empathy remains central even as digital interaction becomes more prevalent.

While Digital Empathy among humans is still under study, Digital Empathy between humans and digital environments has already forged ahead, particularly in how digital environments project empathy toward humans. Picard (1995) pioneered the field of affective computing, emphasizing the critical importance of enabling computers to recognize and respond to human emotions effectively. This technological advancement aimed to humanize digital interactions, ensuring that devices and systems can understand and appropriately react to emotional cues. Picard further discussed the potential for computers to express and even possess emotions, which she argued could enhance their decision-making capabilities and improve their utility in assisting humans. Building on this concept, Kozak (2005) explored the development of digital entities capable of mimicking human empathy, proposing a new frontier where digital agents possess emotional capabilities. This evolution mirrors the empathetic relationships between humans and digital beings outlined in our definition of Digital Empathy, underscoring the profound shift in how empathy operates in digital environments.

Both the "Joy of Fish" story and the German term "Einfühlung," exploring empathy between humans and nature rather than solely between humans as we commonly use it today, serve as a model for developing Digital Empathy to foster meaningful interactions with the increasingly digital environments we inhabit. Just as Zhuangzi's 'Joy of Fish' captured empathy beyond human relationships, Digital Empathy invites us to broaden our understanding of emotional connections within digital environments. Will we dare to explore the "Joy of Robert," a modern reflection of how we empathize with digital entities, like AI agents or Digital Humans, in digital environments, just as Zhuangzi did with fish in natural environments?

The ongoing evolution of Digital Empathy is closely linked to the broader societal and technological shifts. As communication increasingly moves digital, the ability to express and interpret empathy digitally becomes more critical. McDuff and Czerwinski (2020) discuss the potential of technologies to create empathetic user experiences by accurately sensing and responding to user emotions in real time. Digital Empathy will continue to evolve, as our definition indicates. It prompts a continuous reevaluation of the role of technology in our social lives, influencing how we connect with and understand each other across digital mediums. The challenge moving forward will be to ensure that these technologies enhance rather than replace the human elements of empathy, maintaining a balance that respects privacy and enhances interpersonal connections.

**Reflecting Guiding Principles**

Digital Empathy is not just a thought-provoking concept worth further exploration but a vibrant thread that weaves through each of the guiding principles, enhancing and being enhanced by each:

- Centering Humanistic Values: Digital Empathy realizes our commitment to humanistic values by ensuring that digital solutions are designed with a deep understanding of the emotional and psychological needs of users. It empowers older adults to navigate digital platforms that respect their dignity and enhance their quality of life. Furthermore, it enables other members of the society to better understand aging and act accordingly.

- Embracing Digital: Digital Empathy guides the development of intuitive and accessible interfaces and tools for digital technologies, reducing the digital divide and promoting inclusivity among all age groups. For instance, empathetic chatbots for healthcare can check in on older adults daily, reminding them of medications or appointments while enabling them to engage confidently with the digital world.
- Cultivating Learning Societies: Digital Empathy enriches educational content and transforms pedagogical approaches. It enables educators to connect meaningfully with learners, tailoring content to meet diverse emotional and cognitive landscapes. By integrating empathetic technologies, an AI-driven learning platform designed for older adult learners could monitor their packing and comfort level, offering gentle encouragement and simplified explanations when needed, promoting lifelong learning without intimidation.
- Advancing Inclusiveness: Digital Empathy plays a vital role in advancing inclusiveness, ensuring that digital innovations cater to a broad spectrum of abilities and preferences, fostering environments where all individuals feel valued and understood. Voice-activated smart assistants can be tailored for older users with hearing impairments, employing empathetic language prompts and clear speech to make everyday interactions more intuitive and less isolating.
- Taking Holistic Approaches: Digital Empathy enables us to take a holistic approach to education and interaction, considering the full spectrum of user experiences in our design and implementation strategies. In a smart home learning ecosystem, multiple AI-driven devices—such as virtual tutors, home sensors, and medication trackers—communicate with each other empathetically, sharing user mood indicators or emotional cues. By coordinating their responses, these "digital beings" create a more supportive, seamless environment that addresses the older adult's social, cognitive, and physical needs in unison.
- Encouraging Global Knowledge Sharing: Digital Empathy enhances our ability to share knowledge globally. Cyber-Seniors Virtual Programs (n.d.) was initially designed to foster empathetic intergenerational interaction, but now it also brings together older adults from around the world. By sharing their expertise and life experiences within an environment founded on the human-to-human aspect of Digital Empathy, participants foster deeper cross-cultural understanding, bridge both generational and geographical gaps, and enrich mutual learning on a global scale.
- Fostering Adaptability: Digital Empathy supports our principle of adaptability, ensuring that as technologies and user needs evolve, our approaches remain responsive and considerate of the human element at every turn. In a smart home environment, wearables, telepresence robots, and scheduling AIs "talk" to one another about changes in an older user's behavior—such as reduced mobility or increased stress signals. By sharing these insights empathetically, the digital beings dynamically change interface complexity or offer timely interventions (e.g., medication reminders, movement prompts), maintaining the user's comfort and dignity.

Digital Empathy not only enlightens our understanding of human interactions with digital environments but also lays the foundation for our Learning Ecosystem in the digital era, which we will explore in the next section. By grounding each guiding principle in practical examples, we see how Digital Empathy profoundly benefits older adults—while simultaneously improving the overall user experience. It ensures that as we navigate the complexities of digital aging, our technologies, and methodologies remain deeply rooted in understanding and respecting the human condition. By prioritizing empathy in our increasingly digitized world, we enhance the functionality of our digital tools and, more importantly, the lives of those they are designed to serve.

## Learning Ecosystem

The learning ecosystem of "Learning Societies for Digital Aging" is founded on the seven guiding principles that shape the way individuals, communities, and societies approach and navigate the digital world in the context of aging. We will explore how these empathetic interactions underpin the structured approaches of formal learning, the flexibility of nonformal learning, the organic nature of informal learning, and the innovation driven by new learning strategies—all crucial for supporting an aging population in a digital world. Each of these components, while distinct in their structure and delivery, collectively contributes to developing comprehensive, inclusive, and adaptive learning environments that support and enable older adults and societies to thrive in the digital age.

### Empowering the Learning Ecosystem

For our proposed Learning Ecosystem that will be discussed in the next section, Digital Empathy plays a pivotal role in each component:

- Formal Learning: Guides the development of empathetic curricula that respect and respond to the varied needs of learners, particularly older adults.
- Nonformal Learning: Shapes programs that engage community members in meaningful ways, acknowledging their experiences and fostering connections that enhance learning.
- Informal Learning: Supports environments where individuals feel free to explore and express themselves, facilitating natural and effective learning interactions.
- Innovative Strategies: Drives the adoption of advanced technologies that can detect and adapt to the emotional states of users, making learning more dynamic and personalized.

**Formal Learning**

Formal Learning typically occurs within the structured environment of educational institutions and leads to recognized qualifications or certifications. In the context of digital aging, formal Learning plays a crucial role in preparing younger generations with the knowledge and skills necessary to develop technologies, policies, and other measures to support aging populations in a digital world. By integrating digital aging concepts into formal learning settings, we can foster a more inclusive digital society that caters to the needs of older adults. This section will discuss three main aspects of formal Learning: academic education, professional training, and online learning.

***Academic Education***

School, college, and university programs that lead to degrees or certifications provide students with a structured learning environment and a solid foundation in their chosen field of study. In the context of digital aging, academic education can integrate digital aging concepts and equip young people with the knowledge and skills to support older adults in the digital world. In June 2022, the Social Research Center for Welfare Technology at the University of Eastern Finland started to offer a credited graduate course consisting of a webinar series (I-CAFÉ, 2023) covering a variety of topics, including aging scenarios, digital skill learning for older adults, unmet social and healthcare needs in Finland, and the development of a gerontological social work practice pedagogical program in China.

***Professional Training***

Vocational training, industry certifications, and continuing education programs for professional development help individuals acquire the necessary skills to thrive in their careers. By incorporating digital aging components, professional training can prepare young professionals to develop and implement effective strategies, policies, and technologies that address the unique needs of older adults in a digital society. From 2018 to 2021, the School of Social Development and Public Policy at Fudan

University held an annual graduate summer school titled "Age-Friendly Technology and Social Development: A Multidisciplinary Perspective." This program brought together experts from diverse fields to form a multidisciplinary teaching team, including medical anthropology, sociology, gerontology, social work, social policy, demography, information science theory, and medical humanities. More than 200 students from various backgrounds and industries attended the summer school, including graduate and undergraduate students, as well as professionals in social work, eldercare service design, and eldercare product engineering. Combining this multidisciplinary approach with the "service-learning" teaching method enabled young learners to gain multifaced insights into digital aging.

***Online Learning***

Accredited online courses and degree programs that provide formal education through digital platforms offer flexible and accessible learning opportunities for individuals of all ages. By including digital aging topics in online learning curricula, we can ensure that young learners understand the challenges and opportunities associated with an aging population in the digital world.

***Case Study: The Smart Eldercare Undergraduate Course at The Hefei University of Technology, China***

The Smart Eldercare Course at the Hefei University of Technology, China, represents a pioneering effort in formal learning for digital aging. From the central committee to local governments, the Chinese government has developed policies in recent years to support the Smart Eldercare service industry (Chen et al., 2022). This undergraduate course responds to these policies and aims to prepare students for the challenges and opportunities of the growing eldercare market. With its unique blend of online and offline teaching methods, multidisciplinary approach, and focus on practical applications, the Smart Eldercare Course offers an innovative academic model for addressing the complex needs of digital aging.

Innovative Teaching Methods: The course has been offered for six semesters since 2018, with more than 1,600 students enrolled so far. In response to the COVID-19 pandemic, the course has adapted to various teaching formats, including entirely online and blended online-offline approaches. The course leverages the ready online resources, including over 131 minutes of video materials and real-life case studies, to foster a deeper appreciation of smart eldercare.

Multidisciplinary Approach: Recognizing that eldercare requires a combination of skills and knowledge from various disciplines, the Smart Eldercare Course is designed for students from all majors at the university. Students are encouraged to work together in mixed teams to solve practical problems using their diverse backgrounds and expertise. This multidisciplinary approach fosters innovation and prepares students for the collaborative nature of real-world eldercare solutions.

International Collaboration: The course actively engages in international online teaching collaborations, enabling students to learn from global experts in the field. These collaborations have included joint online teaching activities with the Danish company Nordic Praxis on training caregivers with virtual reality technology and exploring Danish welfare technology advancements through virtual visits to the Guldmann (Guldmann, 2023) showroom.

Cultural and Ethical Values: Besides its technical focus, the Smart Eldercare Course aims to promote appreciating the cultural and ethical values of eldercare. The course emphasizes the importance of respecting and honoring older adults, drawing on the rich heritage of traditional Chinese culture. This focus on ethics and values helps students develop a more holistic understanding of eldercare and fosters a compassionate approach to addressing the needs of digital aging.

In summary, the Smart Eldercare Course at the Hefei University of Technology offers a unique and innovative model for formal learning in digital aging. By combining cutting-edge teaching methods, a multidisciplinary approach, international collaboration, and a focus on cultural values, this course has prepared students to become the next generation of leaders in the rapidly evolving field of digital aging.

**Nonformal Learning**

Nonformal learning opportunities typically occur outside the formal education system and provide more flexible, accessible, and diverse learning experiences for older adults. Nonformal learning enables individuals to develop new skills, knowledge, and competencies tailored to their unique needs and interests, fostering lifelong learning and personal growth. In the context of digital aging, nonformal learning can help bridge the digital divide, empower older adults to adapt to the rapidly changing digital landscape, and enhance their overall well-being. This section will explore three components of nonformal Learning: lifelong learning initiatives, community-based programs, and self-paced online courses.

***Lifelong Learning Initiatives***

Lifelong learning initiatives are programs and activities designed to promote continuous learning and personal development throughout an individual's life. These initiatives can range from leisure and hobby-based activities to more structured skill-building courses. For older adults in the digital age, lifelong learning initiatives can help maintain cognitive function, support social engagement, and facilitate acquiring new skills and knowledge relevant to their evolving needs and interests. Additionally, these initiatives can contribute to the overall well-being and quality of life for older adults as they navigate the digital world. Van Vlaardingen (2023) reported the news of Miriam Tees, a 100-year-old woman who continues to attend classes at McGill University's Community for Lifelong Learning. This inspiring example demonstrates how a renowned higher education institution nurtures a learning society for older adults by offering informal, accessible, and engaging learning opportunities that cater to their interests and needs.

***Community-Based Programs***

These programs include workshops, seminars, and courses organized by community centers or local organizations, specifically tailored to the needs and interests of the community members. In the context of digital aging, community-based programs can help older adults enhance their digital skills and knowledge, foster social connections, and provide a supportive learning environment. Such programs can also raise awareness about the potential benefits and challenges of digital technology for aging populations. In addition, community-based programs can go online, form digital communities, and offer virtual programs.

***Self-Paced Online Courses***

Self-paced online courses offer structured learning opportunities that can be completed at the learner's own pace without formal accreditation. These courses provide flexibility and autonomy, allowing older adults to learn new digital skills or deepen their understanding of digital technologies in a manner that suits their individual needs and schedules. Self-paced online courses can cover a wide range of topics, from basic computer literacy to more advanced digital competencies, providing a valuable resource for older adults seeking to enhance their digital proficiency and adapt to the changing digital landscape.

***Case Study: Cyber-Seniors Virtual Programs***

The Cyber-Seniors program （n.d.） exemplifies nonformal Learning in action. Founded in 2015 by the creators of the award-winning documentary film CYBER-SENIORS (Furlong, 1995), this non-profit organization empowers older adults through technology training using an intergenerational volunteer model. Cyber-Seniors leverages the tech-savviness of younger generations to educate older adults in digital technology. The program provides young volunteers with lessons and learning activities to prepare them as digital mentors, while senior citizens gain access to practical technology training and intergenerational communities that keep them socially connected and engaged.

The Cyber-Seniors program belongs to nonformal learning as it is organized and structured, yet it occurs outside the traditional educational system. It is designed to empower older adults through technology training and provides learning opportunities for both younger and older participants.

Furthermore, the Cyber-Seniors program also leveraged intergenerational learning and a volunteer model. This method promotes the exchange of knowledge and skills between different generations and fosters mutual understanding and respect.

## Informal Learning

Informal learning plays a unique role in the learning ecosystem of "Learning Societies for Digital Learning" as it encompasses various learning opportunities that occur naturally in our everyday lives. Informal learning is often self-directed and unstructured, and it can take the following forms within the learning ecosystem.

### *Family-Based Learning*

Family-based Learning is another vital component of informal learning in the digital context. Learning from family members, often through shared experiences, daily activities, or storytelling, can help bridge the generational digital divide. By sharing digital skills and knowledge across generations, families can support each other in navigating the digital world, fostering a sense of belonging and understanding.

### *Peer-to-Peer Learning*

In the digital age, peer-to-peer learning can take various forms, such as exchanging tips on using smartphones, tablets, or other devices, sharing experiences with online platforms and applications, or providing support for troubleshooting digital problems. This learning process can occur in person, through casual conversations or organized gatherings, or virtually via online forums, social media, or video calls.

Peer-to-peer learning is particularly beneficial for older adults as it leverages their social networks and relationships, helping to reduce feelings of isolation or loneliness that can accompany the digital divide. Additionally, learning from peers can help build confidence in using digital technologies, as older adults may feel more comfortable asking questions or seeking guidance from individuals with similar backgrounds or experiences. Overall, peer-to-peer learning for digital aging can enhance the well-being and digital literacy of older adults, empowering them to connect with an increasingly digital world.

### ***Self-Directed Exploration***

Self-directed exploration in the context of Digital Aging highlights the autonomy of older adults and promotes lifelong learning. It can take various forms, such as browsing the internet for information on specific topics, experimenting with digital devices or software, participating in online communities or forums, or even creating and sharing digital content. By engaging in self-directed exploration, older adults can expand their digital horizons, develop new skills, and stay connected with the world around them.

This type of learning is particularly beneficial for older adults as it enables them to take control of their digital learning journey, allowing them to focus on areas of personal interest and relevance. This autonomy fosters a sense of ownership and self-efficacy, which can increase motivation and confidence in using digital technologies. Moreover, self-directed exploration can help older adults remain intellectually engaged and mentally active, contributing to their overall well-being and cognitive health.

### ***Case Study: Self-Directed Exploration at the Apple Store***

Mr. Huang, a 70-year-old retiree, is an inspiring example of self-directed exploration in informal learning for digital aging (Jiang, 2020). Huang first discovered the courses offered by the Apple store in Shanghai in September 2019, and within three months, he had already attended 20 classes, covering topics from primary device usage to photography, drawing, programming, and music editing.

With the primary focus on iPhone and iPad courses, Huang meticulously took notes during each class and revisited the material to deepen his understanding. Despite the repetition of certain content, he found value in reinforcing his learning through multiple exposures to the same course materials. By engaging in this self-directed exploration, Huang has become adept at using the basic functions of his iPhone and iPad, expanding his digital skills and staying connected in the digital world.

Huang's informal learning journey was not limited to Apple courses. He initially started attending photography classes at Sony. His persistence and motivation to learn were evident in his decision to upgrade his devices and invest in his own learning materials.

In addition to attending classes, Huang diligently organized and reviewed his notes to ensure he retained the information. He also sought practical applications for his new skills, such as transferring his scanned family photos to the cloud to preserve them for future generations.

Mr. Huang's experience demonstrates the value of informal learning and self-directed exploration in digital aging. By taking the initiative to attend free courses and immerse himself in learning new technologies, Huang has not only acquired valuable digital skills but also serves as an inspiring example for older adults looking to stay connected and engaged in today's digital world.

**Innovative Learning Strategies**

Innovative learning strategies are instrumental in the development of learning societies for digital aging by bridging the gaps between formal, nonformal, and informal learning. These strategies aim to create a cohesive learning experience, enabling individuals of all ages to acquire and enhance the knowledge, skills, and competencies needed to thrive in a rapidly evolving digital world.

By leveraging cutting-edge pedagogies, methods, and technologies, innovative learning strategies can break down traditional barriers to learning and foster a more inclusive and accessible learning environment. Such strategies include intergenerational learning, blended learning, gamification, adaptive learning systems, virtual or augmented reality, virtual visits, and collaborative online platforms, which can all contribute to a more engaging and personalized learning experience.

Furthermore, innovative learning strategies can facilitate the integration of learning experiences across various settings and contexts, promoting a culture of lifelong learning and continuous development. This approach encourages learners to actively participate in their learning journey, cultivating a sense of ownership, self-efficacy, and resilience.

Using innovative learning strategies to weave formal, nonformal, and informal learning into different local ecosystems can form robust learning societies. These learning societies, in turn, can help ensure that individuals from all walks of life have the opportunity to learn and grow throughout their lives, ultimately enhancing their well-being, social connectedness, and ability to adapt to the ever-changing digital landscape. By forming the learning ecosystems, these strategies can pave the way for a more resilient, inclusive, and adaptive learning society that benefits everyone.

***Intergenerational Learning***

Intergenerational learning, as grounded in Erik Erikson's life span approach, highlights the potential benefits of interactions between individuals at different stages of their lives. Erikson's work demonstrates that the young and old have parallel developmental needs, which can result in a unique synergy between these generations (Erikson, 1963). This approach connects younger and older generations, fostering knowledge exchange and promoting mutual understanding. By leveraging the strengths and experiences of both age groups, intergenerational learning encourages collaboration and helps bridge the digital divide. For example, the Cyber-Seniors program, discussed earlier, utilizes the tech-savviness of younger individuals to educate older adults in digital technology while fostering intergenerational communities that keep older adults socially connected and engaged.

***Collaborative Learning***

Emphasizing the notion that everyone can contribute to knowledge sharing on digital aging, collaborative learning creates a dynamic environment where individuals work together and draw upon each other's expertise. This strategy encourages diverse perspectives and experiences, leading to innovative solutions and approaches to address digital aging challenges. Collaborative learning can occur in various settings, including college classrooms, community workshops, online forums, Wiki sites, or social media platforms, where people of different ages and backgrounds can engage and contribute to the learning process.

***Immersive Learning***

Leveraging technologies such as virtual reality (VR) and augmented reality (AR), immersive learning provides realistic and engaging experiences for learners, enhancing their understanding of complex concepts or situations. In the context of digital aging, immersive learning can be particularly beneficial for training caregivers, healthcare professionals, and older adults themselves. For example, VR simulations can help caregivers gain a deeper understanding of the challenges faced by older adults with cognitive or physical impairments, while AR applications can assist older adults in learning new skills or navigating their environments more effectively. This innovative learning strategy offers a unique opportunity to develop empathy and practical skills to support the aging population in the digital era.

***Case Study: Immersive Learning***

Embodied Labs, founded by Carrie Shaw, is a prime example of immersive learning in the context of digital aging (Sahota, 2020). Shaw was inspired to create a virtual reality (VR) solution to help caregivers better understand the aging community after struggling to explain her mother's early-onset Alzheimer's disease and visual impairment to her caregivers.

Shaw's journey began with a simple invention: a pair of glasses designed to help explain her mother's disease. She then turned to VR to create a more comprehensive and immersive solution that would allow caregivers to step into the shoes of their elderly patients, experiencing the challenges that come with age and disease.

Embodied Labs, headquartered in Los Angeles, California, brings together developers, medical experts, and seasoned Hollywood filmmakers to create realistic, immersive experiences for caregivers and students. The VR programs take participants on a perceptual journey, experiencing life from another person's perspective. Current modules include the Alfred Lab, which follows an African-American man with age-related macular degeneration; the Beatriz Lab, which follows a Latina's 10-year journey through the three stages of Alzheimer's disease; and the Clay Lab, which embodies a Vietnam veteran diagnosed with terminal cancer who, with his family and hospice workers present, experiences the end of life.

Through VR headsets, participants enter the world of their elderly patients, gaining a deeper understanding of their experiences and needs. The training program is for home care, senior living, hospice, medical and nursing schools, hospitals, and other employers. One compelling module is the end-of-life experience, which allows participants to walk through the perspective of someone diagnosed with terminal cancer and decide how they want to live out the end of their life. This immersive learning experience is precious for those entering hospice care as a career, providing an understanding of an end-of-life experience they may not have encountered before.

Embodied Labs' innovative approach to training caregivers has already impacted hundreds of professionals and students across America. The immersive learning platform helps bridge the gap between caregivers and the elderly community, enabling better care and empathy for those affected by age-related diseases and conditions. By leveraging the power of VR, Embodied Labs demonstrates how innovative learning strategies can play a crucial role in addressing the challenges of digital aging and fostering more effective, compassionate care.

## Conclusion

This paper has outlined a learning ecosystem of Learning Societies for Digital Aging that integrates formal, nonformal, and informal learning, as well as innovative learning strategies to support the evolving needs of an aging population in the digital world. The guiding principles of centering humanistic values, embracing digital, cultivating learning societies, advancing inclusiveness, taking holistic approaches, encouraging global knowledge sharing, and fostering adaptability serve as the foundation for building learning societies that empower individuals across the lifespan.

The digital explosion presents opportunities and challenges for the aging population, necessitating the creation of inclusive learning environments that encourage lifelong learning and facilitate the acquisition of new knowledge, skills, and competencies. By integrating digital aging concepts into academic education, professional training, and online learning, we can prepare younger generations to develop technologies, policies, and other measures that cater to the unique needs of the aging population in digital societies.

Nonformal and informal learning settings, such as lifelong learning initiatives, community-based programs, self-paced online courses, peer-to-peer learning, and self-directed exploration, provide valuable opportunities for older adults to stay engaged, connected, and informed. Innovative learning strategies, including intergenerational, collaborative, and immersive learning, could bridge the gap between formal, nonformal, and informal learning, creating cohesive and supportive ecosystems for the aging population.

To nurture learning societies for digital aging, fostering global networks and partnerships, sharing knowledge and resources, and focusing on adaptability to meet the evolving needs of aging populations worldwide is essential. By adopting a holistic and inclusive approach, we can ensure that individuals at all stages of life, especially late stage, have the opportunity to thrive in an ever-changing digital landscape.